\documentclass[twocolumn,secnumarabic,amssymb, nobibnotes, aps, prd]{revtex4-1}

\usepackage{amsmath}

\begin{document}

\title{Uniqueness of the fractional derivative definition}%

\author{Richard Herrmann}%
\email[email:]{herrmann@gigahedron.de}
\affiliation{Berliner Ring 80, D-63303 Dreieich, Germany}
\date{\today}%
\begin{abstract}
For the Riesz fractional derivative besides the well known integral representation two new differential representations are presented, which emphasize the local aspects of a fractional derivative. The consequences for a valid solution of the fractional Schr\"odinger equation are discussed. 
\end{abstract}

\maketitle
\tableofcontents

\section{Introduction}
The concept of action-at-a-distance has dominated the interpretation of physical dynamic behavior in the early years of classical mechanics at the times of Kepler \cite{kep09} and Newton \cite{new92}. In the second half of the 19th century the introduction of a field, first successfully applied in the electromagnetic theory of Maxwell \cite{max73},  has led to a change of paradigm away from the previously  accepted non-local view to an emphasis of local aspects of a given interaction. This development  found its culmination in  the case of gravitational interaction with Einstein's  geometric interpretation in terms of a space deformation \cite{adl75}.

%On the other hand, the success of quantum theory may be interpreted as a renaissance of non-local concepts in physics. Furthermore the interest in non-local field theories is steadily increasing. 
On the other hand, the increasing success of quantum theory since the beginning of the 20th century may be interpreted as a renaissance of non-local concepts in physics and at the time present the interest in non-local field theories is steadily growing. 

A non-relativistic description of quantum particles may be  given using the Schr\"odinger wave equation. In this case, a new facet of non-locality may be introduced extending  the standard local Laplace operator by its non-local fractional pendant. Fractional calculus introduces the concept of non-locality to arbitrary hitherto local operators 
\cite{old76}-\cite{her11}. This is a new property, which  only  recently attracted attention on a broader basis.

Let the one dimensional fractional stationary Schr\"odinger equation in scaled canonical form be defined as 
\begin{equation}
-\Delta^{\alpha/2} \Psi(x) = (E - V(x)) \Psi(x)
\end{equation}
with the fractional Laplace-operator $\Delta^{\alpha/2}$. 
The definition of a fractional order derivative is not unique,  several definitions 
e.g. the Riemann\cite{rie47}, Caputo\cite{cap67},   Riesz  \cite{riesz}, Feller\cite{feller}   coexist and are equally well suited for an extension of the standard derivative.

In order to preserve  Hermiticity  for the fractional extension of the Laplace-operator \cite{laskin}, we will explicitly consider the Riesz fractional derivative \cite{riesz},\cite{he111}:
\begin{eqnarray}
\label{q12driesz}
\Delta^{\alpha/2} f(x)  &\equiv&   _\textrm{\tiny{RZ}}^\infty \partial_x^\alpha f(x)  \\
&=&  \Gamma(1+\alpha)
{\sin(\pi \alpha/2)\over \pi} \times \nonumber  \\
&&
 \int_0^\infty \! 
{f(x+\xi)-2 f(x) + f(x-\xi) \over \xi^{\alpha+1}}d\xi \nonumber \\
& &  \qquad   \qquad \qquad  \qquad  \qquad  0< \alpha <2 \nonumber
\end{eqnarray}
where the left superscript in $_\textrm{\tiny{RZ}}^\infty \partial_x^\alpha$ emphasizes the fact, that the integral domain is the full space $\mathbb{R}$ and therefore explicitly denotes the non-local aspect of this definition.

Since there is an actual discussion about non-local aspects of a fractional derivative \cite{laskin}-\cite{luc13}, in the following we will investigate the uniqueness of its definition. Within our discourse, we will present equivalent local representations of the Riesz derivative, which may be considered as a legitimation of e.g. piecewise solution of a fractional wave equation.
%------------------------------------------
\begin{table}[t]
\caption{coefficients $a_n$ for a central difference approximation of the second derivative of the form $\sum_{n=-N}^N a_n f(x + n \xi)$, from \cite{for88} used in the definition of the generalized Riesz derivative }
\label{tab2ndorderderivative} 
\begin{tabular}{r|rrrrrrrrr}
\hline\noalign{\smallskip}
$N$      &  $a_0$ & $a_{\pm1}$ & $a_{\pm2}$  & $a_{\pm3}$& $a_{\pm4}$ & $a_{\pm5}$ & $a_{\pm6}$ & $a_{\pm7}$   \\
\noalign{\smallskip}\hline\noalign{\smallskip}
$1$&  $-2$               & $1$                                      &  & & & && \\ 
$2$& $-\frac{5}{2}$ &$\frac{4}{3}$ &$-\frac{1}{12}$     & & & &&  \\ 
$3$&  $-\frac{49}{18}$ &$\frac{3}{2}$ &$-\frac{3}{20}$ &$\frac{1}{90}$& & & &  \\ 
$4$& $-\frac{205}{72}$ &$\frac{8}{5}$ &$-\frac{1}{5}$ &$\frac{8}{315}$ &$-\frac{1}{560}$  & & &  \\ 
$5$&$-\frac{5269}{1800}$ &$\frac{5}{3}$ &$-\frac{5}{21}$ &$\frac{5}{126}$ &$-\frac{5}{1008}$ &$\frac{1}{3150}$ & &  \\
$6$&$-\frac{5369}{1800}$ &$\frac{12}{7}$ &$-\frac{15}{56}$ &$\frac{10}{189}$ &$-\frac{1}{112}$ &$\frac{2}{1925}$ &$-\frac{1}{16632}$ &  \\
$7$& $-\frac{266681}{88200}$ &$\frac{7}{4}$ &$-\frac{7}{24}$ &$\frac{7}{108}$ &$-\frac{7}{528}$ &$\frac{7}{3300}$ 
&$-\frac{7}{30888}$ &$\frac{1}{84084}$  \\
\noalign{\smallskip}\hline\noalign{\smallskip}
\end{tabular}
\end{table}

\section{Uniqueness on a global scale - the integral representation}
In order to investigate the uniqueness of a definition of a fractional derivative we will consider the Riesz derivative as an example. In this section, we will investigate a set of generalized  integral representations.

We may determine the Riesz derivative as a specific symmetrized non-local generalization of the standard second order derivative \cite{he111}. We have  
\begin{eqnarray}
_\textrm{\tiny{RZ}}^\infty\partial_x^\alpha  f(x)  &=& I^\alpha  \partial_x^2 f(x) \\
  &=& 
2 \Gamma(\alpha-1) \frac{\sin(\pi \alpha/2)}{\pi}    \times \\
  & &   \int_{0}^{\infty} \!\!\!  d\xi \, \xi^{1-\alpha} {f^{''}(x+\xi)+f^{''}(x-\xi) \over 2} \nonumber\\
&=&
2 \Gamma(\alpha) \frac{\sin(\pi \alpha/2)}{\pi}   \times  \\
& &
     \int_{0}^{\infty} \!\!\!  d\xi \, \xi^{1-\alpha} {f^{'}(x+\xi)-f^{'}(x-\xi) \over 2 \xi} \nonumber\\
\label{wcentrald}
 &=&
\Gamma(1+\alpha) \frac{\sin(\pi \alpha/2)}{\pi}    \times  \\
& &
     \int_{0}^{\infty} \!\!\!  d\xi \, \xi^{1-\alpha} {f(x+\xi)- 2 f(x) + f(x-\xi) \over \xi^2}  \nonumber
\end{eqnarray} 
which indeed looks as a unique definition for the Riesz fractional derivative. 
The eigenfunctions of this operator are the trigonometric functions and the eigenvalues are given by:
\begin{eqnarray} 
\label{origRiesz1}
_\textrm{\tiny{RZ}}^\infty\partial_x^\alpha \exp(i k x) &=& -|k|^\alpha \exp(i k x) \\
\label{c105cos}
_\textrm{\tiny{RZ}}^\infty\partial_x^\alpha \cos(k x) &=& -|k|^\alpha \cos(k x)\\ 
\label{origRiesz3}
_\textrm{\tiny{RZ}}^\infty\partial_x^\alpha \sin(k x) &=& -|k|^\alpha \sin(k x) 
\end{eqnarray} 
Since (\ref{wcentrald}) is nothing else but a weighted sum of the simplest central difference approximation of the second derivative:
\begin{equation}
f^{''}(x)=  {f(x+\xi)- 2 f(x) + f(x-\xi) \over \xi^2} + o\big(f^{(4)}(\xi)\big)
\end{equation}
we may consider more sophisticated  definitions of the Riesz derivative as a result of using higher accuracy approximations of the standard second order derivative, which are given as a finite series over $2 N + 1$ elements
\begin{equation}
f^{''}(x)=  \frac{1}{\xi^2} \sum_{n=-N}^N a_n  f(x+n \xi) + o\big(f^{(2 N + 2)}(\xi)\big)
\end{equation}
with the properties, resulting from the requirement of a vanishing second derivative for a constant function and invariance under parity transformation $\Pi(\pm \xi)$ with positive parity:
\begin{equation}
\label{symmcd}
\sum_{n=-N}^N a_n=0, \quad a_{-n}= a_n
\end{equation}
In table \ref{tab2ndorderderivative} we have compiled the lowest representations of these finite series for $N=1,..,7$.

Therefore we may define the following generalization of the Riesz fractional derivative $_\textrm{\tiny{N}}^\infty\partial_x^\alpha$:
\begin{eqnarray}
\label{rieszgeneralized}
_\textrm{\tiny{N}}^\infty\partial_x^\alpha  f(x)  &=& 
\Gamma(1+\alpha) \frac{\sin(\pi \alpha/2)}{\pi}   \times \\
&&
     \int_{0}^{\infty} \!\!\!  d\xi \, \xi^{1-\alpha} \!\! \sum_{n=-N}^N a_n  f(x+n \xi) \frac{1}{\xi^2} \nonumber
\end{eqnarray} 
which at a first glance looks like a new family of fractional derivatives.

We will choose a pragmatic point of view and will investigate the eigenvalue spectrum of this set of operators. For that purpose, we use the following  properties of the trigonometric functions \cite{Ab}:
\begin{eqnarray}
\label{trigadds}
\sin(z_1 \pm z_2) 
  &=&  \sin(z_1) \cos(z_2) \pm \cos(z_1) \sin(z_2)\\
\label{trigaddc}
\cos(z_1 \pm z_2) 
  &=&  \cos(z_1) \cos(z_2) \mp \sin(z_1) \sin(z_2)
\end{eqnarray} 
For $f(x) = \cos(k x)$ it follows with (\ref{symmcd})
\begin{eqnarray}
\sum_{n=-N}^N && \!\!\!\!\!\!\!a_n \cos(k (x + n \xi)) \nonumber \\
&=& a_0 \cos(k x) + \\
&& \sum_{n=1}^N a_n ( \cos(k (x - n \xi))+ \cos(k (x + n \xi))) \nonumber \\
&=& \cos(k x) \big( a_0  + 2 \sum_{n=1}^N a_n  \cos (k n \xi) \big)
 \end{eqnarray} 
as a consequence, $\cos(k x)$ is an eigenfunction of the generalized Riesz derivative operator $_\textrm{\tiny{N}}^\infty\partial_x^\alpha$. The same statement also holds for $\sin(k x)$.
It follows:
\begin{eqnarray}
_\textrm{\tiny{N}}^\infty\partial_x^\alpha  \exp(i k x) &=& \kappa \exp(i k x)\\
_\textrm{\tiny{N}}^\infty\partial_x^\alpha  \cos(k x) &=& \kappa \cos(k x)\\
_\textrm{\tiny{N}}^\infty\partial_x^\alpha  \sin(k x) &=& \kappa \sin(k x)
 \end{eqnarray} 
with the eigenvalue spectrum $\kappa$:
\begin{eqnarray}
\kappa  &=& \Gamma(1+\alpha) \frac{\sin(\pi \alpha/2)}{\pi} \times \\   
   &&  \int_{0}^{\infty} \!\!\!  d\xi \, \xi^{-1-\alpha} \, \big( a_0  + 2 \sum_{n=1}^N a_n  \cos (k n \xi) \big)
\end{eqnarray} 
For $n=1$ we obtain the Riesz result $\kappa = -|k|^\alpha$. 

For $n>1$ since the integral covers the whole $\mathbb{R}^+$,   we may apply a coordinate transformation of the type $\nu \xi = \hat{\xi}$ to each term in the sum above. It then  follows:
\begin{eqnarray}
\kappa  & =& 
\Gamma(1+\alpha) \frac{\sin(\pi \alpha/2)}{\pi}  \times \\ 
 &&    \int_{0}^{\infty} \!\!\!  d\xi \, \xi^{-1-\alpha} \, \big( a_0  +    2 \sum_{n=1}^N  n^\alpha a_n  \cos (k \xi) \big)\nonumber \\
&=& \Gamma(1+\alpha) \frac{\sin(\pi \alpha/2)}{\pi}  \times \\ 
 &&   |k|^\alpha \int_{0}^{\infty} \!\!\!  d\xi \, \xi^{-1-\alpha} \, \big( a_0  +    2 \sum_{n=1}^N  n^\alpha a_n  \cos (\xi) \big)\nonumber \\
&=& |k|^\alpha  ( a_0  +   \sum_{n=1}^N  n^\alpha a_n ), \quad \quad 0 < \alpha < 2
 \end{eqnarray} 
or in short hand notation:
\begin{equation}
\kappa  = -\zeta_0(\alpha,N) |k|^\alpha
\end{equation} 
Up to a scaling constant $\zeta_0(\alpha, N)=-(a_0 +  \sum_{n=1}^N  n^\alpha a_n )$ the eigenvalue spectrum is identical with the original Riesz derivative eigenvalue spectrum (\ref{origRiesz1})-(\ref{origRiesz3}) and may be absorbed by proper normalization of the generalized Riesz derivative definition. 

Therefore all generalized derivative definitions of type (\ref{rieszgeneralized}), which obey conditions (\ref{symmcd}) are equivalent and lead to same results. In that sense, the Riesz definition of a second order fractional derivative is indeed unique
and emphasizes the non-local aspects of a fractional derivative.

It should be emphasized, that alternative realizations of the fractional Riesz derivative in terms of e.g. a central differences representation of Gr\"unwald-Letnikov type are equivalent to the above integral representation in their non-local behavior\cite{ort06}-\cite{ort12}.

In the next section we will investigate the differential representation of the Riesz derivative and will demonstrate in a similar way as in the case of the integral representations that different approaches lead to the same result.
\section{Uniqueness on a local scale - the differential representation}
Differential representations of the Riemann and Caputo fractional derivative in  terms of a  series expansion of integer derivatives are commonly known \cite{sam93,her11,tar07,tar08b}. This approach emphasizes the local aspects of a fractional derivative. A corresponding series expansion for the Riesz derivative would lead to a local representation of the same fractional derivative.

Indeed there are several strategies to derive a differential representation of the Riesz derivative. Let us begin with the fractional extension of the binomial series \cite{Ab},\cite{liu10}:
\begin{equation}
\label{binseries1}
\partial_x^\alpha  = \lim_{\omega \rightarrow 0} (\partial_x + \omega)^\alpha 
= \lim_{\omega \rightarrow 0} \sum_{j=0}^\infty 
\binom{\alpha}{j}\,
\omega^{\alpha-j} {\partial_x^j} \quad \alpha \in \mathbb{R}
\end{equation}  
where  $\omega$ is an arbitrary real number. 

Motivated by  the correspondence
\begin{equation}
\label{RZlim}
\lim_{\alpha \rightarrow 2} {_\textrm{\tiny{RZ}}}\partial_x^\alpha  =  \partial_x^2
\end{equation}  
which holds for the Riesz derivative, we extend the above binomial series  to
\begin{eqnarray}
\label{binseries2}
_{\tiny{2}}^{\tiny{\triangle}}\partial_x^\alpha  &=& \lim_{\omega \rightarrow 0} (\partial_x^2 + \omega^2)^{\alpha/2} \\
&=&\lim_{\omega \rightarrow 0}  \sum_{j=0}^\infty 
\binom{\alpha/2}{j}\,
(\omega^2)^{\alpha/2-j} {\partial_x^{2 j}} \\                          
&=&\lim_{\omega \rightarrow 0}  |\omega|^\alpha  \sum_{j=0}^\infty 
\binom{\alpha/2}{j}\,
|\omega|^{-2 j} {\partial_x^{2 j}} \quad  \alpha \in \mathbb{R}
\end{eqnarray} 
where the superscript $_{\tiny{2}}^{\tiny{\triangle}}$ 
emphasizes the differential representation of an hitherto integral  representation of a fractional derivative operator $\partial_x^\alpha$. 

Applying this operator to the exponential function leads to:
\begin{eqnarray}
\label{binseries2}
_{\tiny{2}}^{\tiny{\triangle}}\partial_x^\alpha  && \!\!\!\!\!\exp(k x)  
\nonumber \\
&=&
\lim_{\omega \rightarrow 0}    \sum_{j=0}^\infty 
\binom{\alpha/2}{j}\,
(\omega^2)^{\alpha/2-j} {k^{2 j}}\exp(k x)  \\                          
&=&\lim_{\omega \rightarrow 0} (k^2 + \omega^2)^{\alpha/2} \exp(k x)  \\  
 &=& |k|^\alpha \exp(k x)  \quad\quad\quad\quad\quad\quad\quad \alpha \in \mathbb{R}
\end{eqnarray} 
Consequently we interpret this operator as the hyperbolic Riesz derivative, since it works for $k \in \mathbb{R}$, while the original Riesz derivative in its integral form is divergent for $\exp(k x)$ but converges for $\exp(i k x)$.

Therefore in an heuristic approach we obtain as a differential representation of the Riesz derivative:
\begin{eqnarray}
_\textrm{\tiny{RZ}}^{\tiny{\triangle}}\partial_x^\alpha                          
&=&\lim_{\omega \rightarrow 0}  -|\omega|^\alpha  \sum_{j=0}^\infty 
\binom{\alpha/2}{j}\,
|\omega|^{-2 j} {(i \partial_x)^{2 j}}\\
\label{dRZ1}
&=&\lim_{\omega \rightarrow 0}  -|\omega|^\alpha  \sum_{j=0}^\infty 
\binom{\alpha/2}{j}\,
|\omega|^{-2 j} (-1)^j {\partial_x^{2 j}} 
\end{eqnarray} 
which we call a valid realization of the differential form of the Riesz fractional derivative.

Indeed it follows in accordance with (\ref{origRiesz1})-(\ref{origRiesz3}):
\begin{eqnarray}
\label{dRiesz1}
_\textrm{\tiny{RZ}}^{\tiny{\triangle}}\partial_x^\alpha \exp(i k x) &=& -|k|^a \exp(i k x) \\
_\textrm{\tiny{RZ}}^{\tiny{\triangle}}\partial_x^\alpha \cos(k x) &=& -|k|^a \cos(k x) \\
\label{dRiesz3}
_\textrm{\tiny{RZ}}^{\tiny{\triangle}}\partial_x^\alpha \sin(k x) &=& -|k|^a \sin(k x)  \qquad \alpha \in \mathbb{R}
\end{eqnarray}
Since we have realized the differential form of the Riesz derivative as the limit of a series we will answer the question if other series may yield the same result.

As a demonstration, we use the fractional extension of the Leibniz product rule \cite{wat31}, \cite{osl70}
\begin{equation}
\label{leibseries1}
\partial_x^\alpha (u(x) v(x))   = 
\sum_{j=0}^\infty 
\binom{\alpha}{j}\,
(\partial_x^{\alpha-j}u(x)) (\partial_x^j v(x)) 
\end{equation} 
and rewrite the analytic function $f(x)$ as
\begin{equation}
 f(x)   = \lim_{\omega \rightarrow 0} \cos(\omega x) f(x) 
\end{equation} 
With (\ref{c105cos})  follows:
\begin{eqnarray}
_\textrm{\tiny{RZ}}\partial_x^{\alpha-j} &&\!\!\!\!\!\!\lim_{\omega \rightarrow 0}  \cos(\omega x) \nonumber \\
            &=& \lim_{\omega \rightarrow 0} (\partial_x^{-j} {_\textrm{\tiny{RZ}}}\partial_x^{\alpha})  \cos(\omega x)\\
            &=&\lim_{\omega \rightarrow 0}  -|\omega|^\alpha \partial_x^{-j} \cos(\omega x) \\
            &=&\lim_{\omega \rightarrow 0}  -|\omega|^\alpha \omega^{-j} \cos(\omega x - \frac{\pi}{2}j)\\
            &=&\lim_{\omega \rightarrow 0}  -|\omega|^\alpha \omega^{-j} \cos(- \frac{\pi}{2}j)\\
            &=&\lim_{\omega \rightarrow 0}  -|\omega|^\alpha \omega^{-j} 
\begin{cases}
0         &  \text{$j$ odd}\cr
   (-1)^{j/2} &   \text{$j$ even}
\end{cases}
\end{eqnarray} 
Using the Leibniz product rule we obtain:
\begin{eqnarray}
\label{leibseries1}
_\textrm{\tiny{2}}^{\tiny{\triangle}}\partial_x^\alpha    &=& \lim_{\omega \rightarrow 0}
\sum_{j=0}^\infty 
\binom{\alpha}{j}\,
(\partial_x^{\alpha-j} \cos(\omega x)) \partial_x^j \\
&=& -\lim_{\omega \rightarrow 0}
|\omega|^\alpha \sum_{j=0}^\infty 
\binom{\alpha}{2 j}\,
\omega^{-2 j} (-1)^j \partial_x^{2 j}  \\
 &=&-\lim_{\omega \rightarrow 0}{|\omega|^\alpha} {_2F_1}(\frac{1}{2}-\frac{\alpha}{2},-\frac{\alpha}{2};\frac{1}{2};-\frac{1}{\omega^2}\partial_x^2)
\end{eqnarray} 
Lets apply this operator to the exponential function:
\begin{eqnarray}
_\textrm{\tiny{2}}^{\tiny{\triangle}}\partial_x^\alpha && \!\!\!\!\!\exp(k x)   \nonumber \\
  &=& 
-\lim_{\omega \rightarrow 0}{|\omega|^\alpha} {_2F_1}(\frac{1}{2}-\frac{\alpha}{2},-\frac{\alpha}{2};\frac{1}{2};-\frac{k^2}{\omega^2}) \exp(k x) \nonumber \\
&=& -|k|^\alpha \cos(\alpha \pi/2) \exp(k x)
\end{eqnarray}
where we have used (15.3.7) and (15.1.8) from \cite{Ab}.

Once again,  we may consider this operator as an alternative realization of a hyperbolic Riesz derivative, since it works for $k \in \mathbb{R}$, while the original Riesz derivative in its integral form is divergent for $\exp(k x)$ but converges for $\exp(i k x)$. 

Therefore we define a differential representation of the Riesz derivative heuristically:
\begin{eqnarray}
_\textrm{\tiny{RZ}}^{\tiny{\triangle}}&& \!\!\!\!\!\!\partial_x^\alpha   \nonumber \\
  &=&-\frac{1}{\cos(\alpha \pi/2)}  \lim_{\omega \rightarrow 0}
|\omega|^\alpha \sum_{j=0}^\infty 
\binom{\alpha}{2 j}\,
\omega^{-2 j}  \partial_x^{2 j}  \\
\label{dRZ2}
 &=&-\frac{1}{\cos(\alpha \pi/2)}\lim_{\omega \rightarrow 0}{|\omega|^\alpha} {_2F_1}(\frac{1}{2}-\frac{\alpha}{2},-\frac{\alpha}{2};\frac{1}{2};\frac{1}{\omega^2}\partial_x^2) \nonumber \\
&&
\end{eqnarray} 
with the same eigenfunctions and eigenvalue spectrum (\ref{dRiesz1})-(\ref{dRiesz3}).

Hence we have demonstrated, that indeed there exist distinct differential representations, which in the limit $\omega \rightarrow 0$ lead to the same eigenvalue spectrum. 

\section{Manifest covariant local representation of the Riesz derivative on $\mathbb{R}^N$}
A straightforward extension of e.g. (\ref{dRZ2}) to the N-dimensional case is  given by
\begin{equation}
_\textrm{\tiny{RZ}}^{\tiny{\triangle}}\bigtriangleup^{\alpha/2}_N   =
\label{dRZN}
-\frac{1}{\cos(\alpha \pi/2)}\lim_{\omega \rightarrow 0}{|\omega|^\alpha} {_2F_1}(\frac{1}{2}-\frac{\alpha}{2},-\frac{\alpha}{2};\frac{1}{2};\frac{\bigtriangleup}{\omega^2}) 
\end{equation} 
with the N-dimensional Laplace-operator $\bigtriangleup$ in carthesian coordinates on $\mathbb{R}^N$
\begin{equation}
\bigtriangleup  = \sum_{n=1}^N \partial^2 _{x_n}  \quad N \in \mathbb{N}
\end{equation} 
The eigenfunctions and the eigenvalue spectrum of this differential representation of the  N-dimensional Riesz derivative is then given by:
\begin{eqnarray}
_\textrm{\tiny{RZ}}^{\tiny{\triangle}}\bigtriangleup^{\alpha/2}_N  && \prod_{n=1}^N \exp(i k_n x_n) = \nonumber \\
&&
-(\sum_{n=1}^N k_n^2)^{\alpha/2} \, \prod_{n=1}^N \exp(i k_n x_n)
\end{eqnarray} 
which coincides with the standard result using the standard integral representation.

In a similar approach we may extend (\ref{dRZ1}) to the N-dimensional case:
\begin{eqnarray}
_\textrm{\tiny{RZ}}^{\tiny{\triangle}}\bigtriangleup^{\alpha/2}_N 
\label{dRZM}
&=&\lim_{\omega \rightarrow 0}  -|\omega|^\alpha  \sum_{j=0}^\infty 
\binom{\alpha/2}{j}\,
|\omega|^{-2 j} (-1)^j {\bigtriangleup^{j}} 
\end{eqnarray} 
Since the Laplace-operator may be derived  for any set of coordinates, where the metric tensor $g_{ij}$ is known \cite{adl75} via: 
%-----------------------------------------------------------------------
\begin{eqnarray}
\bigtriangleup        & = &g^{ij} \nabla_{i} \nabla_{j}
      \qquad \qquad i,j=1,...,N                                      \\
  & = &\frac{1}{\sqrt{g}} \partial_{i}\,
        g^{ij} \sqrt{g} \, \partial_{j} \\
        & = & g^{ij}(\partial_{i} \partial_{j} -
% ssssss s t a r t christoffelsymbol  a,b,c
\left\{\begin{array}{c}
k      \\ ij           \\
\end{array} \right\}
% sssssss  e n d   christoffelsymbol  a,b,c
                 \partial_{k})                                         
\end{eqnarray}
%-----------------------------------------------------------------------
where  $\nabla_{i}$ denotes the Riemann covariant derivative, $g$ is the determinant of the metric tensor, $g= \det g_{ij}$
 and 
$\left\{\begin{array}{c}
k      \\ ij           \\
\end{array} \right\}$ is the Christoffel symbol,
(\ref{dRZN}) and (\ref{dRZM}) may be considered as two optional candidates for a  differential representation of a valid covariant realization of  the Riesz fractional derivative on the Riemannian space.   
 
\section{Conclusion}
With (\ref{dRZN}) and (\ref{dRZM}) we have presented two different realizations of a differential representation of the Riesz derivative as a limiting case of two different series expansions in terms of integer derivatives.

At least in the case of the trigonometric functions $\sin(k x)$ and $\cos(k x)$ and therefore for every Fourier series  these series are convergent and valid for all $\alpha \in \mathbb{R}$ and thus are more robust than their integral counterparts, where the range of allowed $\alpha$ values is restricted to $0 < \alpha < 2$.

It is important to mention, that these representations are realized as series in terms of standard derivatives and therefore determine a local version of the fractional derivative, since information is required only within an $\epsilon$-region around $x$. 
The use of a  fractional derivative does not automatically imply non-locality.

As a consequence, using the differential representations of the Riesz derivative, it seems a valid procedure to generate piecewise steady solutions of a Riesz type  fractional Schr\"odinger equation even though in a way  this contradicts Feynman's view of a path integral formulation of  quantum mechanics.  

Consequently while in standard quantum mechanics Schr\"odinger's wave equation as a local view and Feynman's path integral approach as a non-local view lead to equivalent results, in fractional quantum mechanics this equivalence obviously is lost and leads to different results \cite{laskin}-\cite{luc13}. 

Hence using the Riesz fractional derivative given in terms of  either the integral or the differential representation indeed makes a difference in e.g. fractional wave equations. Emphasizing fundamentally different aspects of a local or non-local approach to physical problems the use of the Riesz fractional derivative  in either form revives the discussion of concepts like action-at-a-distance.

\begin{acknowledgments}
We thank A. Friedrich and M.
Ortigueira, FCT, Lisboa, Portugal for useful discussions.
\end{acknowledgments}


\end{document}